\documentclass[12pt]{article}
\usepackage{rotating}
\usepackage{setspace}
\usepackage{epsfig}
\usepackage{amssymb}
\def\be{\begin{equation}}
\def\ee{\end{equation}}
\def\ba{\begin{array}}
\def\ea{\end{array}}
\def\beqn{\begin{eqnarray}}
\def\eeqn{\end{eqnarray}}
\def\nonum{\nonumber}
\def\bt{\begin{tabular}}
\def\et{\end{tabular}}
\def\bc{\begin{center}}
\def\ec{\end{center}}
\begin{document}
\title{$~~~~~~~~$Texture 4 zero Fritzsch-like lepton mass matrices}
 \author{Gulsheen Ahuja$^{1}$,
 Sanjeev Kumar$^{2}$,
Monika Randhawa$^{3}$,\\
 Manmohan Gupta$^{1,}$\thanks{mmgupta@pu.ac.in},
 S. Dev$^{2}$\\
\\{$^1$ \it Department of Physics, Centre of Advanced Study, P.U.,
 Chandigarh, India.}\\{$^2$ \it Department of Physics, H.P.U., Shimla,
 India.}\\{$^3$ \it University Institute of Engineering and Technology, P.U., Chandigarh,
 India.} }
 \maketitle
 \begin{abstract}
For Majorana or Dirac neutrinos, using Fritzsch-like texture 4
zero mass matrices with parallel texture structures for the
charged leptons and the Dirac neutrino mass matrix ($M_{\nu D}$),
detailed predictions for cases pertaining to normal/inverted
hierarchy as well as degenerate scenarios of neutrino masses have
been carried out. The inverted hierarchy as well as degenerate
scenarios seem to be ruled out at $3 \sigma$ C.L. for both
Majorana and Dirac neutrinos. For normal hierarchy, Jarlskog's
rephasing invariant parameter $J$, the CP violating Dirac-like
phase $\delta$ and the effective neutrino mass $ \langle m_{ee}
\rangle$ have been calculated. For this case, lower limits of
$m_{\nu_1}$ and $\theta_{13}$ would have implications for the
nature of neutrinos.
 \end{abstract}
In the last few years, impressive advances have been made in
understanding the phenomenology of neutrino oscillations through
solar neutrino experiments \cite{solexp}, atmospheric neutrino
experiments \cite{atmexp}, reactor based experiments
\cite{reacexp} and accelerator based experiments \cite{accexp}. At
present, one of the key issues in the context of neutrino
oscillation phenomenology is to understand the pattern of neutrino
masses and mixings which seems to be vastly different from that of
quark masses and mixings. In the case of quarks the masses and
mixing angles show distinct hierarchy, whereas in the case of
neutrinos neither the mixing angles nor the neutrino masses show
any distinct hierarchy. In fact, the two mixing angles governing
solar and atmospheric neutrino oscillations look to be rather
large, the third angle may be very small compared to these.
Further, at present there is no consensus about neutrino masses
which may show normal/inverted hierarchy or may even be
degenerate. Furthermore, the situation becomes complicated when
one realizes that neutrino masses are much smaller than charged
lepton and quark masses.

Taking clue from the success of texture specific mass matrices in
the context of quarks \cite{quarkstex}, several attempts
\cite{neuttex, leptex} have been made to consider similar lepton
mass matrices for explaining the pattern of neutrino masses and
mixings. In the absence of sufficient amount of data regarding
neutrino masses and mixing angles, it would require a very careful
scrutiny of all possible textures to find viable structures which
are compatible with data and theoretical ideas so that these be
kept in mind while formulating mass matrices at the GUT (Grand
Unified Theories) scale. In this context, most of the attempts to
understand the pattern of neutrino masses and mixings have been
carried out using the seesaw mechanism \cite{seesaw} given by
 \be M_{\nu}=-M_{\nu D}^T\,(M_R)^{-1}\,M_{\nu D},
\label{seesaweq} \ee \noindent where $M_{\nu D}$ and $M_R$ are
respectively the Dirac neutrino mass matrix and the right-handed
Majorana neutrino mass matrix. In this context, it should be noted
that the predictions are somewhat different, on the one hand when
texture is imposed only on $M_{\nu D}$ and $M_R$ and on the other
hand when $M_{\nu}$ and $M_{\nu D}$ have the same texture by
imposing `texture invariant conditions' \cite{xingtexinv,matsuda}.
Similarly, as compared to the Majorana case, the predictions are
different for Dirac neutrinos even if $M_{\nu D}$ has the same
texture as that of $M_{\nu}$.

Texture 4 zero mass matrices are known to explain the pattern of
quark masses and mixings \cite{t4quarkstex} as well as are known
to be compatible with specific models of GUTs \cite{quarkstex,
neuttex} and these could be obtained using considerations of
Abelian family symmetries \cite{grimus}. Assuming normal hierarchy
of masses as well as imposing texture 4 zero structure on $M_{\nu
D}$ and charged lepton mass matrices, Xing {\it et al.}
\cite{xingtexinv} have not only shown the compatibility of these
with neutrino oscillation phenomenology but have also shown the
seesaw invariance of these structures under certain conditions.
Very recently, Matsuda {\it et al.} \cite{matsuda} have reiterated
the success of texture 4 zero mass matrices in the case of quarks
by showing that these mass matrices can accommodate the present
value of sin$2\beta$. Also, for normal hierarchy they have shown
that texture 4 zero lepton mass matrices can accommodate large
values of mixing angle $s_{13}$. In particular, by imposing
texture invariant conditions they have shown that $M_{\nu}$ can be
texture 2 zero when one assumes Fritzsch-like texture 2 zero
structure for $M_{\nu D}$, $M_R$ as well as for charged lepton
mass matrix.

In view of the success of texture 4 zero mass matrices in the
context of quark mixing phenomenology, it would be interesting to
carry out an exhaustive and detailed analysis of these in the case
of leptons. In particular, for both Majorana and  Dirac neutrinos
it would be interesting to investigate the compatibility of
texture 4 zero lepton mass matrices with the inverted hierarchy
and degenerate scenario of neutrino masses. In the case of
Majorana neutrinos due to the absence of any guidelines for $M_R$,
to keep the number of independent parameters under control, it
would perhaps be desirable to keep its structure as simple as
possible. It would also be desirable to study the implications of
these mass matrices when texture is imposed only on $M_{\nu D}$
and not on $M_{\nu}$.

Considering Fritzsch-like texture 4 zero lepton mass matrices,
with neutrinos and charged leptons having parallel texture
structures, the purpose of the present communication is to
investigate in detail the implications of these for
normal/inverted hierarchy as well as `different' degenerate
scenarios of neutrino masses. In particular, for the inverted
hierarchy and degenerate scenarios, the implications of these
structures have been carried out for both Majorana and Dirac
neutrinos by exploring the parameter space available to any of the
two mixing angles found by giving full variation to other
parameters and phases. In the case of normal hierarchy, we have
calculated several phenomenological quantities such as Jarlskog's
rephasing invariant parameter $J$, the CP violating Dirac-like
phase $\delta$ and the effective neutrino mass $ \langle m_{ee}
\rangle$, related to neutrinoless double beta decay
$(\beta\beta)_{0 \, \nu}$.

To begin with, we define the modified Fritzsch-like matrices,
e.g.,
 \be
 M_{l}=\left( \ba{ccc}
0 & A _{l} & 0      \\ A_{l}^{*} & D_{l} &  B_{l}     \\
 0 &     B_{l}^{*}  &  C_{l} \ea \right), \qquad
M_{\nu D}=\left( \ba{ccc} 0 &A _{\nu} & 0      \\ A_{\nu}^{*} &
D_{\nu} &  B_{\nu}     \\
 0 &     B_{\nu}^{*}  &  C_{\nu} \ea \right),
 \label{frzmm}
 \ee
$M_{l}$ and $M_{\nu D}$ respectively corresponding to Dirac-like
charged lepton and neutrino mass matrices. The above matrices are
texture 4 zero type with $D_l$ and $D_{\nu}$ being non-zero along
with $A_{l(\nu)} =|A_{l(\nu)}|e^{i\alpha_{l(\nu)}}$
 and $B_{l(\nu)} = |B_{l(\nu)}|e^{i\beta_{l(\nu)}}$, in case these
 are symmetric then $A_{l(\nu)}^*$ and $B_{l(\nu)}^*$ should be
 replaced by $A_{l(\nu)}$ and $B_{l(\nu)}$, as well as
 $C_{l(\nu)}$ and $D_{l(\nu)}$ should respectively be defined as $C_{l(\nu)}
 =|C_{l(\nu)}|e^{i\gamma_{l(\nu)}}$ and $D_{l(\nu)}
 =|D_{l(\nu)}|e^{i\omega_{l(\nu)}}$.

To fix the notations and conventions as well as to facilitate the
understanding of inverted hierarchy case and its relationship to
the normal hierarchy case, we detail the formalism connecting the
mass matrix to the neutrino mixing matrix. The mass matrices $M_l$
and $M_{\nu D}$ given in equation (\ref{frzmm}), for Hermitian as
well as symmetric case, can be exactly diagonalized. Details of
Hermitian case can be looked up in our earlier work \cite{mon},
the symmetric case can similarly be worked out. To facilitate
diagonalization, the mass matrix $M_k$, where $k=l, \nu D$, can be
expressed as
\be
M_k= Q_k M_k^r P_k \,  \label{mk} \ee or  \be M_k^r= Q_k^{\dagger}
M_k P_k^{\dagger}\,, \label{mkr} \ee where $M_k^r$ is a real
symmetric matrix with real eigenvalues and $Q_k$ and $P_k$ are
diagonal phase matrices. For the hermitian case $Q_k=
P_k^{\dagger}$, whereas for the symmetric case under certain
conditions $Q_k= P_k$. In general, the real matrix $M_k^r$ is
diagonalized by the orthogonal transformation $O_k$, e.g., \be
M_k^{diag}= {O_k}^T M_k^r O_k \,, \label{mkdiag} \ee which on
using equation (\ref{mkr}) can be rewritten as \be M_k^{diag}=
{O_k}^T Q_k^{\dagger} M_k P_k^{\dagger} O_k \,. \label{mkdiag2}
\ee To facilitate the construction of diagonalization
transformations for different hierarchies, we introduce a diagonal
phase matrix $\xi_k$ defined as $ {\rm diag} (1,\,e^{i \pi},\,1)$
for the case of normal hierarchy and as $ {\rm diag} (1,\,e^{i
\pi},\,e^{i \pi})$ for the case of inverted hierarchy. Equation
(\ref{mkdiag2}) can now be written as \be \xi_k M_k^{diag}=
{O_k}^T Q_k^{\dagger} M_k P_k^{\dagger} O_k \,, \label{mkdiag3}
\ee which can also be expressed as \be M_k^{diag}= \xi_k^{\dagger}
{O_k}^T Q_k^{\dagger} M_k P_k^{\dagger} O_k \,. \label{mkdiag4}
\ee Making use of the fact that $O_k^*=O_k$ it can be further
expressed as
\be
M_k^{diag}=(Q_k O_k \xi_k)^{\dagger} M_k (P_k^{\dagger}
O_k),\label{mkeq} \ee from which one gets \be M_k=Q_k O_k \xi_k
M_k^{diag} O_k^T P_k.\label{mkeq2} \ee

The case of leptons is fairly straight forward, for the neutrinos
the diagonalizing transformation is hierarchy specific as well as
requires some fine tuning of the phases of the right handed
neutrino mass matrix $M_R$. To clarify this point further, in
analogy with equation (\ref{mkeq2}), we can express $M_{\nu D}$ as
\be M_{\nu D}=Q_{\nu D} O_{\nu D} \xi_{\nu D} M_{\nu D}^{diag}
O_{\nu D}^T P_{\nu D}.\label{mnud} \ee Substituting the above
value of $M_{\nu D}$ in equation (\ref{seesaweq}) one obtains
\be
M_{\nu}=-(Q_{\nu D} O_{\nu D} \xi_{\nu D} M_{\nu D}^{diag} O_{\nu
D}^T P_{\nu D})^T (M_R)^{-1} (Q_{\nu D} O_{\nu D} \xi_{\nu D}
M_{\nu D}^{diag} O_{\nu D}^T P_{\nu D}). \ee On using $P_{\nu D}^T
= P_{\nu D}$, the above equation can further be written as
\be
M_{\nu}=-P_{\nu D} O_{\nu D} M_{\nu D}^{diag} \xi_{\nu D} O_{\nu
D}^T Q_{\nu D}^T (M_R)^{-1} Q_{\nu D} O_{\nu D} \xi_{\nu D} M_{\nu
D}^{diag} O_{\nu D}^T P_{\nu D}. \ee Assuming fine tuning, the
phase matrices $Q_{\nu D}^T$ and $Q_{\nu D}$ along with $-M_R$ can
be taken as $m_R ~{\rm diag} (1,1,1)$ as well as using the
unitarity of $\xi_{\nu D}$ and orthogonality of $O_{\nu D}$, the
above equation can be expressed as
\be
M_{\nu}= P_{\nu D} O_{\nu D} \frac{(M_{\nu
D}^{diag})^2}{(m_R)^{-1}} O_{\nu D}^T P_{\nu D}. \label{mnu} \ee

The lepton mixing matrix, obtained from the matrices used for
diagonalizing the mass matrices $M_l$ and $M_{\nu}$, is expressed
as
 \be
U =(Q_l O_l \xi_l)^{\dagger} (P_{\nu D} O_{\nu D}). \label{mix}
\ee Eliminating the phase matrix $\xi_l$ by redefinition of the
charged lepton phases, the above equation becomes
\be
 U = O_l^{\dagger} Q_l P_{\nu D} O_{\nu D} \,, \label{mixreal} \ee
where $Q_l P_{\nu D}$, without loss of generality, can be taken as
$(e^{i\phi_1},\,1,\,e^{i\phi_2})$, $\phi_1$ and $\phi_2$ being
related to the phases of mass matrices and can be treated as free
parameters.

To understand the relationship between diagonalizing
transformations for different hierarchies of neutrino masses as
well as their relationship with the charged lepton case, we first
consider the general diagonalizing transformation $O_k$, whose
elements can be written as
 \beqn
O_k(11) & = & {\sqrt \frac{m_{2} m_{3} (m_{3}+m_{2}-D_{l(\nu)})}
     {(m_{1}+m_{2}+m_{3}-D_{l(\nu)})
(m_{1}-m_{3})(m_{1}-m_{2})} } \nonum  \\ O_k(12) & = & {\sqrt
\frac{m_{1} m_{3}
 (m_{1}+m_{3}-D_{l(\nu)})}
   {(m_{1}+m_{2}+m_{3}-D_{l(\nu)})
 (m_{2}-m_{3})(m_{2}-m_{1})} }
\nonum   \\O_k(13) & = & {\sqrt \frac{m_{1} m_{2}
 (m_{2}+m_{1}-D_{l(\nu)})}
    {(m_{1}+m_{2}+m_{3}-D_{l(\nu)})
(m_{3}-m_{2})(m_{3}-m_{1})} } \nonum   \\ O_k(21) & = & {\sqrt
\frac{m_{1}
 (m_{3}+m_{2}-D_{l(\nu)})}
  {(m_{3}-m_{1})(m_{1}-m_{2})} }
\nonum  \\O_k(22) & = & {\sqrt \frac{m_{2}
(m_{3}+m_{1}-D_{l(\nu)})}
  {(m_{2}-m_{3})(m_{2}-m_{1})} }
 \nonum    \\
O_k(23) & = & \sqrt{\frac{m_3(m_{2}+m_{1}-D_{l(\nu)})}
 {(m_{2}-m_{3})(m_{3}-m_{1})} }
\nonum   \\O_k(31) & = &
 \sqrt{\frac{m_{1} (m_{1}+m_{2}-D_{l(\nu)})
    (m_{1}-m_{3})(m_{1}+m_{3}-D_{l(\nu)})}
{(m_{1}-m_{3})^2 (m_{1}-m_{2})(m_{1}+m_{2}+m_{3}-D_{l(\nu)})}}
\nonum
\\O_k(32) & = & {\sqrt \frac{m_{2}(D_{l(\nu)}-m_{1}-m_{2})
(D_{l(\nu)}-m_{2}-m_{3})}{(m_{1}+m_{2}+m_{3}-D_{l(\nu)})
 (m_{2}-m_{3})(m_{2}-m_{1})} }
 \nonum  \\
O_k(33) & = & {\sqrt \frac{m_{3}(D_{l(\nu)}-m_{2}-m_{3})
(D_{l(\nu)}-m_{1}-m_{3})}{(m_{1}+m_{2}+m_{3}-D_{l(\nu)})
 (m_{3}-m_{1})(m_{3}-m_{2})}}\,,  \label{diaggen}
 \eeqn where $m_1$, $m_2$,
$m_3$ are eigenvalues of $M_k$. In the case of charged leptons,
because of the hierarchy $m_e \ll m_{\mu} \ll m_{\tau}$, the mass
 eigenstates can be approximated respectively to the flavor
eigenstates as has been considered by several authors \cite{
neuttex,fuku}. Using the approximation, $m_{l1} \simeq m_e$,
$m_{l2} \simeq m_{\mu}$ and $m_{l3} \simeq m_{\tau}$, the first
element of the matrix $O_l$ can be obtained from the corresponding
element of equation (\ref{diaggen}) by replacing $m_1$, $m_2$,
$m_3$ by $m_e$, $-m_{\mu}$, $m_{\tau}$, e.g.,
 \be  O_l(11) = {\sqrt
\frac{m_{\mu} m_{\tau} (m_{\tau}-m_{\mu}-D_l)}
     {(m_{e}-m_{\mu}+m_{\tau}-D_l)
(m_{\tau}-m_{e})(m_{e}+m_{\mu})} } ~. \ee

For normal hierarchy defined as $m_{\nu_1}<m_{\nu_2}\ll
m_{\nu_3}$, as well as for the corresponding degenerate case given
by $m_{\nu_1} \lesssim m_{\nu_2} \sim m_{\nu_3}$, equation
(\ref{diaggen}) can also be used to obtain the first element of
diagonalizing transformation for Majorana neutrinos. By replacing
$m_1$, $m_2$, $m_3$ by $\sqrt{m_{\nu 1} m_R}$, $-\sqrt{m_{\nu 2}
m_R}$, $\sqrt{m_{\nu 3} m_R}$ in the equation, we get \be
O_{\nu}(11) = {\sqrt \frac{\sqrt{m_{\nu_2}}
    \sqrt{m_{\nu_3}}
( \sqrt{m_{\nu_3}}-\sqrt{ m_{\nu_2}}-D_{\nu})}
{(\sqrt{m_{\nu_1}}-\sqrt{m_{\nu_2}} + \sqrt{m_{\nu_3}}- D_{\nu})
(\sqrt{m_{\nu_3}}-\sqrt{m_{\nu_1}}) (\sqrt{m_{\nu_1}} +
\sqrt{m_{\nu_2}} )} } \label{omajnh}, \ee where $m_{\nu_1}$,
$m_{\nu_2}$ and $m_{\nu_3}$ are neutrino masses. The parameter
$D_{\nu}$ is to be divided by $\sqrt{m_R}$, however as $D_{\nu}$
is arbitrary therefore we retain it as it is.

In the same manner, one can obtain the elements of diagonalizing
transformation for the inverted hierarchy case defined as
$m_{\nu_3} \ll m_{\nu_1} < m_{\nu_2}$ as well as for the
corresponding degenerate case given by $m_{\nu_3} \sim m_{\nu_1}
\lesssim m_{\nu_2}$. By replacing $m_1$, $m_2$, $m_3$ in equation
(\ref{diaggen}) with $\sqrt{m_{\nu_1} m_R}$, $-\sqrt{m_{\nu_2}
m_R}$, $-\sqrt{m_{\nu_3} m_R}$, we obtain \be O_{\nu}(11) = {\sqrt
\frac{\sqrt{m_{\nu_2}}
    \sqrt{m_{\nu_3}}
(D_{\nu}+\sqrt{ m_{\nu_2}} + \sqrt{m_{\nu_3}} )}
{(-\sqrt{m_{\nu_1}}+\sqrt{m_{\nu_2}} + \sqrt{m_{\nu_3}}+ D_{\nu})
(\sqrt{m_{\nu_1}}+\sqrt{m_{\nu_3}}) (\sqrt{m_{\nu_1}} +
\sqrt{m_{\nu_2}} )} } \label{omajih}. \ee The other elements of
diagonalizing transformations in the case of neutrinos as well as
charged leptons can similarly be found. The above formalism has
been presented for Majorana neutrinos, for the Dirac case the
mixing matrix can easily be derived from diagonalizing
transformation of $M_l$ and $M_{\nu D}$.

It may be of interest to mention that in the case of normal
hierarchy, the formulation of Matsuda {\it et al.} \cite{matsuda}
can easily be obtained from the present case. For example, their
element $O_{11}$ can be obtained from our $O_k(11)$ by replacing
`$d_f$' by `$m_1 +m_2 +m_3 + D_{l(\nu)}$', similarly their other
elements can be derived from the elements of our $O_k$. Further,
it should also be noted that they have treated the $3 \times 3$
element of the mass matrix as free parameter whereas in the
present case we have treated $2 \times 2$ element as a free
parameter. Furthermore, we have not put any conditions so as to
obtain a particular texture for $M_{\nu}$.

Before discussing the results, we would like to mention some of
the details pertaining to various inputs. For the purpose of our
analysis we have used the results obtained by a recent global
analysis carried out by Valle \cite{valle}, incorporating solar
\cite{solexp}, atmospheric \cite{atmexp}, reactor \cite{reacexp}
and accelerator based experiments \cite{accexp}. The 3$\sigma$
values of the neutrino mass and mixing parameters so obtained are
\be
 \Delta m_{12}^{2} = (7.1 - 8.9)\times
 10^{-5}~\rm{eV}^{2},~~~~
 \Delta m_{23}^{2} = (2.0 - 3.2)\times 10^{-3}~ \rm{eV}^{2},
 \label{solatmmass}\ee
\be
{\rm sin}^2\,\theta_{12}  =  0.24 - 0.40,~~~
 {\rm sin}^2\,\theta_{23}  =  0.34 - 0.68,~~~
 {\rm sin}^2\,\theta_{13} \leq 0.040. \label{s13}
\ee These values are quite compatible with those obtained very
recently by a global analysis carried out by Garcia and Maltoni
\cite{recgar}. It may be mentioned that the present upper bound on
${\rm sin}\,\theta_{13}$ is somewhat lower than the CHOOZ bound
\cite{chooz}. Further, for the purpose of calculations, we have
taken the lightest neutrino mass, the phases $\phi_1$, $\phi_2$
and $D_{l, \nu}$ as free parameters, the other two masses are
constrained by $\Delta m_{12}^2 = m_{\nu_2}^2 - m_{\nu_1}^2 $ and
$\Delta m_{23}^2 = m_{\nu_3}^2 - m_{\nu_2}^2 $ in the normal
hierarchy case and by $\Delta m_{23}^2 = m_{\nu_2}^2 -
m_{\nu_3}^2$ in the inverted hierarchy case. It may be noted that
lightest neutrino mass corresponds to $m_{\nu_1}$ for the normal
hierarchy case and to $m_{\nu_3}$ for the inverted hierarchy case.
For all the three hierarchies, the explored range of the lightest
neutrino mass is taken to be $10^{-8}\,\rm{eV}-10^{-1}\,\rm{eV}$,
our conclusions remain unaffected even if the range is extended
further. In the absence of any constraint on the phases, $\phi_1$
and $\phi_2$ have been given full variation from 0 to $2\pi$.
Although $D_{l, \nu}$ are free parameters, however, they have been
constrained such that diagonalizing transformations $O_l$ and
$O_{\nu}$ always remain real. This implies, for leptons $-(m_{l_2}
- m_{l_1})<D_{l}< (m_{l_3} + m_{l_2})$, for Dirac neutrinos
$-(m_{\nu_2} - m_{\nu_1}) < D_{\nu}
<(m_{\nu_3} - m_{\nu_2})$ for normal hierarchy and
$-(m_{\nu_2} - m_{\nu_1}) < D_{\nu}
<(m_{\nu_1} - m_{\nu_3})$ for inverted hierarchy. Similarly, for
Majorana neutrinos $-(\sqrt{m_{\nu_2}} -\sqrt{m_{\nu_1}}) <
D_{\nu}
<(\sqrt{m_{\nu_3}} - \sqrt{m_{\nu_2}})$ for normal hierarchy and
$-(\sqrt{m_{\nu_2}} - \sqrt{m_{\nu_1}}) < D_{\nu}
<(\sqrt{m_{\nu_1}} - \sqrt{m_{\nu_3}})$ for inverted hierarchy.
The calculations pertaining to the case when charged leptons are
in flavor basis can easily be deduced from the above calculations
for both Majorana and Dirac neutrinos.

Considering Majorana or Dirac neutrinos, we have carried out
detailed calculations pertaining to texture 4 zero lepton mass
matrices for the possibilities of neutrino masses having
normal/inverted hierarchy or being degenerate. To begin with, we
consider the inverted hierarchy case for both types of neutrinos.
In this context, it may be mentioned that for both the
possibilities texture is imposed only on $M_{\nu D}$, with no such
restriction on $M_{\nu}$ for the Majorana case. In figures 1a, 1b
and 1c, for Majorana neutrinos we have plotted the parameter space
corresponding to any of the two mixing angles by constraining the
third angle by its values given in equation (\ref{s13}) while
giving full allowed variation to other parameters. Also included
in the figures are blank rectangular regions indicating the
experimentally allowed $3\sigma$ region of the plotted angles.
Interestingly, a general look at these figures reveals that the
case of inverted hierarchy seems to be ruled out. A closer look at
these figures brings out several interesting points. From figure
1a showing the plot of angles $\theta_{12}$ versus $\theta_{23}$,
one can immediately conclude that the plotted parameter space
includes the experimentally allowed range of
$\theta_{23}=35.7^{\circ}-55.6^{\circ}$, however it excludes the
experimentally allowed range of
$\theta_{12}=29.3^{\circ}-39.25^{\circ}$. This clearly indicates
that at 3$\sigma$ C.L. inverted hierarchy is not viable. It may be
noted that while plotting this figure, $\theta_{13}$ is restricted
by the bound given in equation (\ref{s13}), while $\Delta
m_{12}^2$ and $\Delta m_{23}^2$ are constrained by the
experimental limits given in equation (\ref{solatmmass}). It may
also be mentioned that although the 3$\sigma$ upper limit of angle
$\theta_{12}$ is not included in the plotted parameter space, yet
it lies very near to the boundary, therefore the above conclusions
needs to be checked further.

The conclusions arrived above can be further checked from figures
1b and 1c wherein we have plotted $\theta_{12}$ versus
$\theta_{13}$ and $\theta_{23}$ versus $\theta_{13}$ respectively
by constraining angles $\theta_{23}$ and $\theta_{12}$. Both the
figures indicate that the plotted parameter space does not include
simultaneously the experimental bounds of the plotted angles,
e.g., $\theta_{12}$ in the case of figure 1b and $\theta_{13}$ in
figure 1c. Here it needs to be mentioned that similar to figure
1a, in figure 1b also the 3$\sigma$ upper limit of angle
$\theta_{12}$ lies very near to the boundary of the plotted
parameter space, however in figure 1c the 3$\sigma$ upper limit of
angle $\theta_{13}$ is well below the plotted parameter space.

For Dirac neutrinos, again inverted hierarchy seems to be ruled
out as can be easily checked from figures 2a, 2b and 2c, plotted
in a manner similar to the Majorana case by constraining one
mixing angle by its experimental limits and plotting the parameter
space for the other two angles. Again, these figures reveal that
the plotted parameter space does not overlap with the experimental
limits of at least one of the plotted angles, thereby indicating
that inverted hierarchy is ruled out at 3$\sigma$ C.L. for Dirac
neutrinos as well.

For Majorana or Dirac neutrinos the case of neutrino masses being
degenerate, characterized by either $m_{\nu_1} \lesssim m_{\nu_2}
\sim m_{\nu_3} \lesssim 0.1~\rm{eV}$ or $m_{\nu_3} \sim m_{\nu_1}
\lesssim m_{\nu_2} \lesssim 0.1~\rm{eV}$ corresponding to normal
hierarchy and inverted hierarchy respectively, is again ruled out.
Considering degenerate scenario corresponding to inverted
hierarchy, figures 1 and 2 can again be used to rule out
degenerate scenario at 3$\sigma$ C.L. for Majorana and Dirac
neutrinos respectively. It needs to be mentioned that while
plotting these figures the range of the lightest neutrino mass is
taken to be $10^{-8}\,\rm{eV}-10^{-1}\,\rm{eV}$, which includes
the neutrino masses corresponding to degenerate scenario,
therefore by discussion similar to the one given for ruling out
inverted hierarchy, degenerate scenario of neutrino masses is
ruled out as well.

Coming to degenerate scenario corresponding to normal hierarchy,
one can easily show that this is ruled out again. To this end, in
figure \ref{th12vm-md}, by giving full variation to other
parameters, we have plotted the mixing angle $\theta_{12}$ against
the lightest neutrino mass $m_{\nu_1}$. Figure \ref{th12vm-md}a
corresponds to the case of Majorana neutrinos and figure
\ref{th12vm-md}b to the case of Dirac neutrinos. From the figures
one can immediately find that the values of $\theta_{12}$
corresponding to $m_{\nu_1} \lesssim 0.1~\rm{eV}$ lie outside the
experimentally allowed range, thereby ruling out degenerate
scenario for Majorana as well as Dirac neutrinos at 3$\sigma$
C.L..

The presence of a few isolated points near the experimentally
allowed 3$\sigma$ regions shown in figures 1 and 2 may raise
doubts about our conclusions. In order to check whether there are
any solution points within the experimentally allowed 3$\sigma$
region of plotted angles, we have attempted to obtain a common
parameter space pertaining to the three mixing angles
simultaneously. Interestingly, we find that all possible cases
considered here pertaining to inverted hierarchy and degenerate
scenario are again ruled out. It may also be added that in the
case when charged leptons are in the flavor basis, one can easily
check that inverted hierarchy and degenerate scenarios for the
texture 4 zero mass matrices are again ruled out, in agreement
with the conclusions of \cite{sanjeev}. The results pertaining to
this case can easily be derived from our earlier cases.

After ruling out the cases pertaining to inverted hierarchy and
degenerate scenario, we now discuss the normal hierarchy cases.
For the charged lepton mass matrix $M_l$ being Fritzsch-like or in
the flavor basis, for Majorana as well as Dirac neutrinos, in
table (\ref{tab1}) we have presented the viable ranges of neutrino
masses, mixing angles $\theta_{12}$, $\theta_{23}$ and
$\theta_{13}$, Jarlskog's rephasing invariant parameter $J$, CP
violating phase $\delta$ and effective neutrino mass $ \langle
m_{ee} \rangle$. The parameter $J$ can be calculated by using its
expression given in \cite{fuku}, whereas $\delta$ can be
determined from $J=s_{12}s_{23}s_{13}c_{12}c_{23}c_{13}^2\, {\rm
sin}\,\delta $ where $c_{ij} = {\rm cos}\theta_{ij}$ and $s_{ij} =
{\rm sin}\theta_{ij}$, for $i,j=1,2,3$. The effective Majorana
mass, related to neutrinoless double beta decay $(\beta\beta)_{0
\, \nu}$, is given as
\be
\langle m_{ee} \rangle = m_{\nu_1} U_{e1}^2 + m_{\nu_2} U_{e2}^2
+m_{\nu_3} U_{e3}^2. \label{mee} \ee

Considering the texture 4 zero case when $M_{\nu D}$ and $M_l$
both have parallel texture structures, a close look at table
(\ref{tab1}) reveals several interesting points. For both Dirac or
Majorana neutrinos, the viable range of the lightest neutrino mass
$m_{\nu_1}$ is quite different, in particular the range
corresponding to Dirac neutrinos is much wider at both the ends as
compared to the Majorana neutrinos. Similar conclusions can be
arrived at by studying the implications of the well known mixing
angle $\theta_{12}$ on the lightest neutrino mass $m_{\nu_1}$
through a closer look at the figures 3a and 3b. Therefore, a
measurement of $m_{\nu_1}$ could have important implications for
the nature of neutrinos. Somewhat constrained range of $m_{\nu_2}$
for the Majorana case as compared to the Dirac case is also due to
the constrained range of $m_{\nu_1}$ for the Majorana case. Also,
from the table one finds that the lower limit on $\theta_{13}$ for
the Dirac case is considerably lower than for the Majorana case,
therefore a measurement of $\theta_{13}$ would have important
implications for this case. It must be noted that the calculated
values of $\langle m_{ee} \rangle$ are much less compared to the
present limits of $\langle m_{ee} \rangle$ \cite{heidel},
therefore, these do not have any implications for the texture 4
zero cases considered here. However, the future experiments with
considerably higher sensitivities, aiming to measure $\langle
m_{ee} \rangle \simeq 3.6\times
 10^{-2}~\rm{eV}$ (MOON \cite{moon}) and $\langle m_{ee} \rangle \simeq 2.7\times
 10^{-2}~\rm{eV}$ (CUORE \cite{cuore}), would have implications on
the cases considered here. The different cases of Dirac and
Majorana neutrinos do not show any divergence for the ranges of
Jarlskog's rephasing invariant parameter .

It may be of interest to construct the
Pontecorvo-Maki-Nakagawa-Sakata (PMNS) mixing matrix \cite{pmns}
which for Majorana neutrinos is
 \be U=\left( \ba{ccc}
 0.7599  -  0.8701   &    0.4797  -  0.6294 &      0.0199  -  0.1994 \\
  0.1673 -  0.5715   &    0.3948  -  0.7606 &      0.5720  -  0.8224 \\
  0.1854  -  0.5912  &    0.3549  -  0.7363 &      0.5540  -
  0.8094
 \ea \right), \label{mmaj} \ee
 wherein we have given the magnitude of the matrix elements.
Similarly, for Dirac neutrinos, the PMNS matrix is
  \be U=\left( \ba{ccc}
 0.7604  -  0.9213  &    0.3887 -  0.6317&      0.0015  -  0.1993 \\
  0.1475  -  0.5552   &    0.4049  -  0.8170 &      0.4154  -  0.8244 \\
  0.1830  -  0.6022   &    0.3648  -  0.7441 &      0.5546  -
  0.9095
 \ea \right). \label{mdir} \ee
A general look at the two matrices reveals that the ranges of the
matrix elements are more wider in the case of Dirac neutrinos as
compared to those in the case of Majorana neutrinos. A comparison
of the two matrices shows that the lower limit of the element
$U_{\mu 3}$ show an appreciable difference, which seems to be due
to the nature of neutrinos, hence, a further precision of $U_{\mu
3}$ would have important implications. Also, it may be mentioned
that both the above mentioned matrices are fully compatible with a
very recent construction of a mixing matrix by Bjorken {\it et
al.} \cite{bjorken} assuming democratic trimaximally mixed $\nu_2$
mass eigenstate as well as with the one presented by Giunti
\cite{giunti}.

For the sake of completion pertaining to normal hierarchy, in
table (\ref{tab1}) we have also presented the results when charged
leptons are in the flavor basis which can be easily deduced from
the case when $M_l$ is Fritzsch-like. Interestingly, from the
table one immediately finds that in this case both $J$ and
$\delta$ are vanishingly small for the wide range of parameters
considered here, which can easily be understood by examining the
corresponding mixing matrix. Also, the range of angle
$\theta_{13}$ is much narrower compared to the case when $M_l$ is
Fritzsch-like, particularly for the Majorana case the predicted
range is very narrow, therefore a measurement of $\theta_{13}$
would have an immediate impact on this case. It may also be added
that for the Majorana case, the range of $\theta_{23}$ is
compatible only with the lower part of the present admissible
range, however for the Dirac case there is no such restriction.
These conclusions are broadly in agreement with those of
\cite{sanjeev}.

To summarize, detailed calculations have been carried out for
different hierarchies in the case of Fritzsch-like texture 4 zero
mass matrices with parallel texture structures for charged leptons
and for Dirac neutrino mass matrix ($M_{\nu D}$) using latest
3$\sigma$ input values of neutrino mass and mixing parameters. For
the inverted hierarchy, pertaining to both Majorana and Dirac
neutrinos, parameter space available to any two of the mixing
angles has been explored while considering wide ranges of free
parameters available. Similarly, the viability of `different'
degenerate scenarios has been examined. Interestingly for both
types of neutrinos, inverted hierarchy as well as degenerate
scenarios seem to be ruled out at $3\sigma$ C.L. and hence
strongly disfavored. It may also be added that the results when
charged leptons are in the flavor basis can easily be deduced from
the present calculations and these lead to similar conclusions.

For the normal hierarchy case, several phenomenological quantities
such as Jarlskog's rephasing invariant parameter $J$, the CP
violating Dirac-like phase $\delta$ and the effective neutrino
mass $ \langle m_{ee} \rangle$ have  been calculated. The
different cases of Majorana and Dirac neutrinos do not show any
divergence for the ranges of $J$ and phase $\delta$. In the case
of $m_{\nu_1}$ and $\theta_{13}$, the Dirac case seems to
accommodate a larger range of these parameters. In particular, a
measurement of the lower limits of these parameters would have
implications for the nature of neutrinos. Also, the PMNS matrices
constructed for Majorana as well as Dirac neutrinos, by giving
full variation to the parameters, are compatible with a very
recent construction of a mixing matrix by Bjorken {\it et al.}
\cite{bjorken} assuming democratic trimaximally mixed $\nu_2$ mass
eigenstate.

 \vskip 0.2cm
{\bf Acknowledgements} \\ M.G. and G.A. would like to thank DAE,
BRNS (grant No.2005/37/4/BRNS), India, for financial support. S.K.
acknowledges the financial support provided by CSIR, India. M.R.
would like to thank the Director, UIET for providing facilities to
work.

\newpage

\begin{table}
\begin{tabular}{|c|cc|cc|} \hline
& \multicolumn{2}{c|} {$M_l$ being Fritzsch-like} &
\multicolumn{2}{c|} {$M_l$ being in the flavor basis} \\ \hline
&Dirac case &Majorana case & Dirac case &Majorana case\\ \hline

$m_{\nu_1}$ & 5.73 $\times 10^{-5}$  - 0.012 & 2.47 $\times
10^{-4}$ - 0.006 & (1.63 - 6.28) $\times 10^{-3}$ & (.402 - 2.06)
$\times 10^{-3}$
\\
$m_{\nu_2}$ & 0.0084 - 0.0149 & 0.0084 - 0.0108 & 0.0086 - 0.0113
& 0.0084 - 0.0096
\\
$m_{\nu_3}$ & 0.0456 - 0.0577 & 0.0455 - 0.0575 & 0.0446 - 0.0576
& 0.0455 - 0.0573
\\
$\theta_{12}$ & 29.30$^{\circ}$ - 39.20$^{\circ}$  &
29.30$^{\circ}$ - 39.20$^{\circ}$ & 29.30$^{\circ}$ -
39.20$^{\circ}$  & 29.30$^{\circ}$ - 39.04$^{\circ}$
\\
$\theta_{23}$ & 35.70$^{\circ}$ - 55.60$^{\circ}$  &
35.70$^{\circ}$ - 55.60$^{\circ}$ & 35.70$^{\circ}$ -
55.59$^{\circ}$  & 35.70$^{\circ}$ - 40.15$^{\circ}$
 \\
$\theta_{13}$ & 0.084$^{\circ}$ - 11.50$^{\circ}$  &
1.14$^{\circ}$ - 11.50$^{\circ}$ & 3.60$^{\circ}$ -
11.15$^{\circ}$ & 8.43$^{\circ}$ - 11.50$^{\circ}$
 \\
$J$ & $-$0.0462 - .0448  & $-$0.0459 - .0463 & $\sim 0$ & $\sim 0$
 \\
$\delta$ & $-90^{\circ}$ - 90.0$^{\circ}$  &$-90^{\circ}$ -
90.0$^{\circ}$ & $\sim 0^{\circ}$  &$\sim 0^{\circ}$
\\
$\langle m_{ee} \rangle$ & - & .00086 - .0173 & - & .0032 - .0075
 \\ \hline
\end{tabular}
 \caption{Calculated ranges for neutrino mass and mixing parameters
obtained by varying $\phi_1$ and $\phi_2$ from 0 to 2$\pi$ for the
normal hierarchy case. Inputs have been defined in the text. All
masses are in $\rm{eV}$.} \label{tab1}
\end{table}

\begin{figure}[hbt]
\vspace{0.12in}
\centerline{\epsfysize=2.8in\epsffile{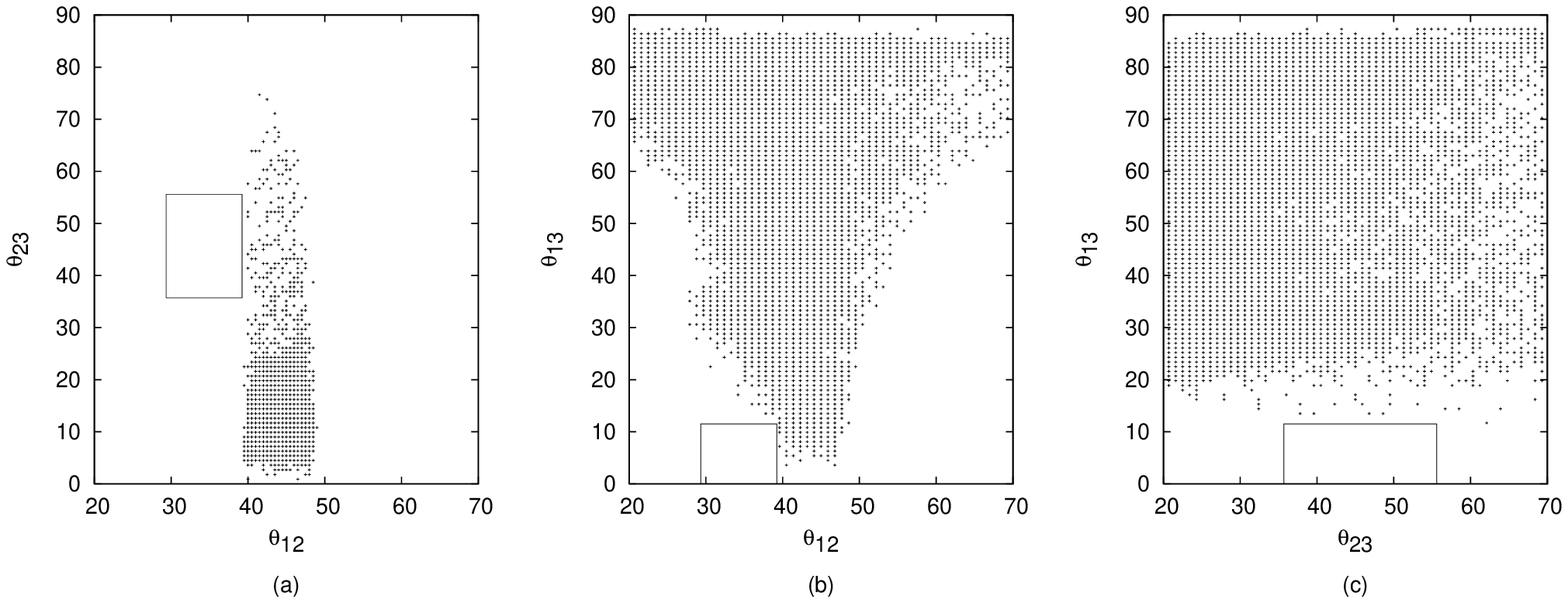}}
\vspace{0.08in}
   \caption{Plots showing the parameter space
corresponding to any of the two mixing angles by constraining the
third angle by its experimental limits given in equation
(\ref{s13}) and giving full allowed variation to other parameters
for Majorana neutrinos. The blank rectangular region indicates the
experimentally allowed $3\sigma$ region of the plotted angles.}
  \label{fig1}
  \end{figure}

  \begin{figure}[tbp]
\centerline{\epsfysize=2.8in\epsffile{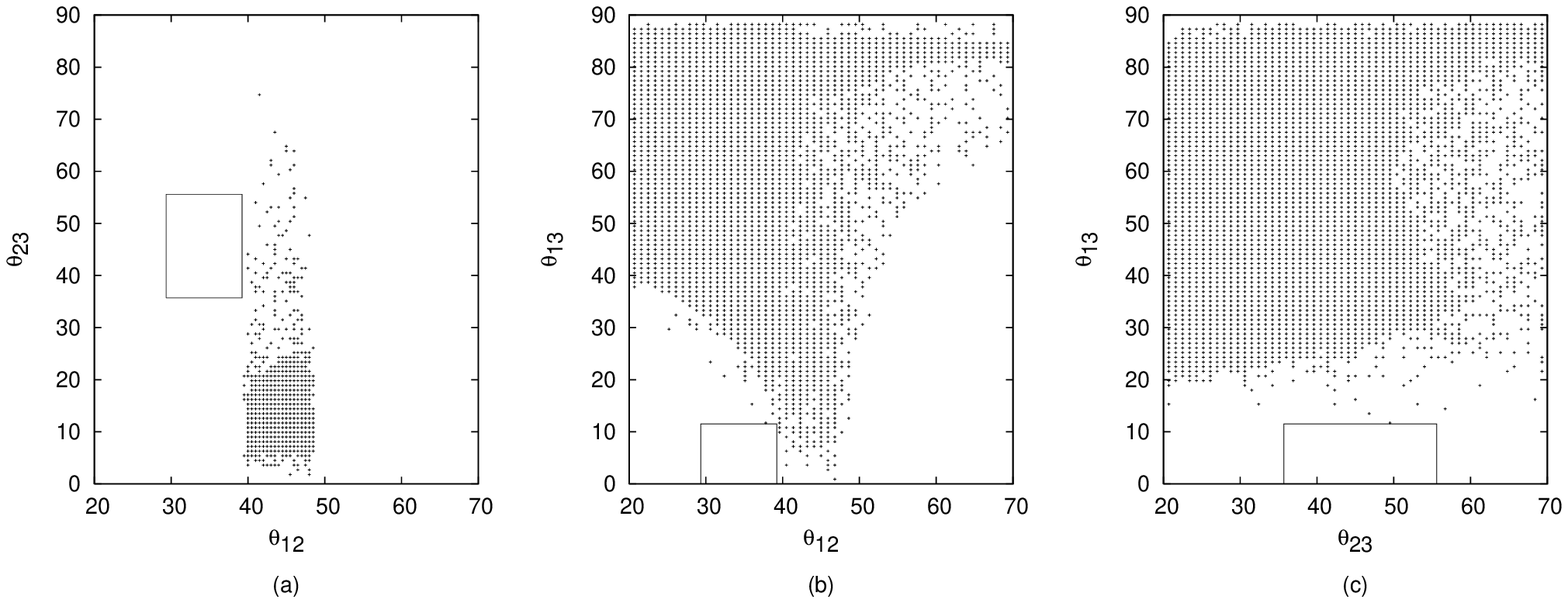}}
\vspace{0.08in}
   \caption{Plots showing the parameter space
corresponding to any of the two mixing angles by constraining the
third angle by its experimental limits given in equation
(\ref{s13}) and giving full allowed variation to other parameters
for Dirac neutrinos. The blank rectangular region indicates the
experimentally allowed $3\sigma$ region of the plotted angles.}
  \label{fig2}
  \end{figure}

  \begin{figure}[tbp]
\centerline{\epsfysize=2.8in\epsffile{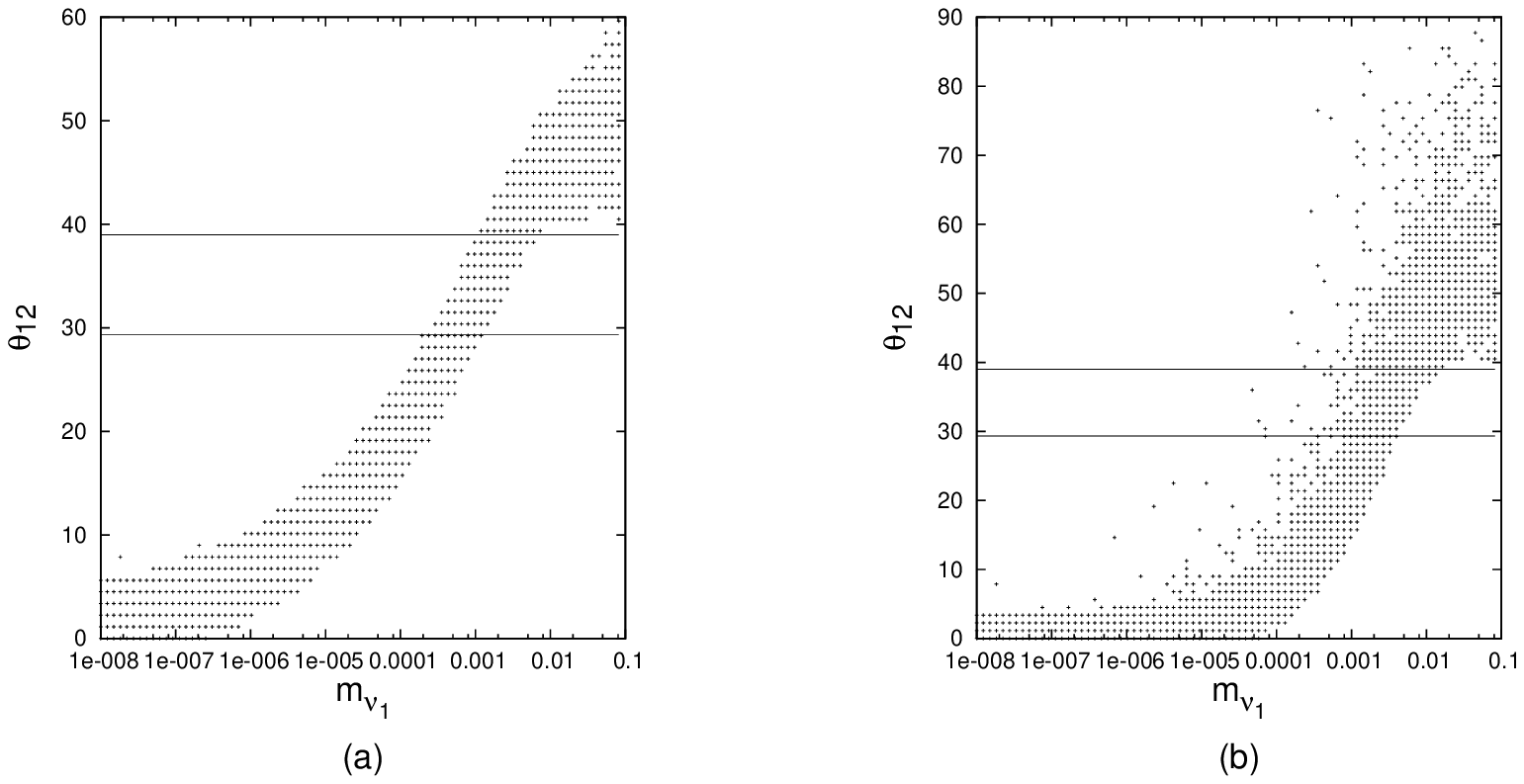}}
\vspace{0.08in}
 \caption{Plots showing variation of mixing angle
$\theta_{12}$ with lightest neutrino mass $m_{\nu_1}$ by giving
full variation to other parameters for neutrinos being (a)
Majorana-like and (b) Dirac-like. The parallel lines indicate the
$3\sigma$ limits of angle $\theta_{12}$.}
  \label{th12vm-md}
  \end{figure}

\end{document}